\documentclass[finals,epj,nopacs,twocolumn]{svjour}

\usepackage{latexsym}
\def\makeheadbox{\relax}

  \newcommand{\Qa}{\mathcal{Q}}
\usepackage{graphics}
\usepackage{hyperref}
\usepackage{mathtools}
\usepackage{amsmath}
\usepackage{cuted}
 \usepackage[latin1]{inputenc}
\usepackage{color}
\usepackage{url}
\usepackage{graphicx}
\usepackage{float}
\usepackage{amsmath}
\usepackage{tikz}
\usepackage{graphicx}
\usepackage{latexsym,color}
\usepackage{amssymb}
\usepackage{xcolor,color}
\usepackage{amsfonts,dsfont}

\usepackage[T1]{fontenc}
\usepackage{ae,aecompl,latexsym}

\definecolor{blue}{rgb}{0.0,0.0,0.75}
\definecolor{purple}{rgb}{0.78,0.0,0.75}
\hypersetup{linkcolor=purple,citecolor=blue,filecolor=cyan,urlcolor=magenta}

\def\be{\begin{equation}}
\def\ee{\end{equation}}
\def\bea{\begin{eqnarray}}

\newcommand{\prd}{Phys.Rev.D}

\newcommand{\aap}{A\&A}

\def\eea{\end{eqnarray}}

\newcommand{\la}{\mathcal{A}}
\newcommand{\laa}{\mathcal{L}}
\usepackage{physics}

\newcommand{\il}{~}

\usepackage[varg]{txfonts}

\begin{document}
\title{Repulsive gravity  effects in horizon formation}
\subtitle{Horizon remnants in  naked singularities}

\author{Daniela  Pugliese\inst{1} \and Hernando Quevedo\inst{2}
}                     
\institute{ $^1$ Research Centre of Theoretical Physics and Astrophysics,
	Institute of Physics,
	Silesian University in Opava,
	Bezru\v{c}ovo n\'{a}m\v{e}st\'{i} 13, CZ-74601 Opava, Czech Republic
	\\     $^{2}$
	Instituto de Ciencias Nucleares, Universidad Nacional Aut\'onoma de M\'exico,  AP 70543, M\'exico, DF 04510, Mexico \\
	Dipartimento di Fisica and ICRA, Universit\`a di Roma ``La Sapienza", I-00185 Roma, Italy \\
	Department of Theoretical and Nuclear Physics,
	Kazakh National University,
	Almaty 050040, Kazakhstan
}
\abstract{Repulsive gravity  is a well known characteristic  of naked singularities. In this work, we explore light surfaces and find  new effects  of repulsive gravity. We compare Kerr naked singularities with the corresponding black hole counterparts and find certain  structures that are identified as horizon remnants. We argue that these features  might be significant for the comprehension of processes that lead to the  formation or eventually destruction of black hole Killing horizons. These features can be detected by observing photon orbits, particularly close to the rotation axis, which can be used to distinguish naked singularities from black holes.
 \keywords{Black holes; naked singularities; Killing horizons;light surfaces; photon orbits.}}
\date{Received: date / Revised version: date}
\maketitle

\section{Introduction}\label{Sec:impe-null-i}
We explore possible effects of repulsive gravity in the formation of  horizons by studying  structures appearing in the light surfaces of a class of very slowing spinning Kerr naked singularities (\textbf{NSs}),  variously studied in literature in connection   with the  Killing horizon  properties of the Kerr black holes (\textbf{BHs})  \cite{bundle-EPJC-complete,observers,remnants,Pugliese:2019efv,Pugliese:2019rfz,LQG-bundles,PS}.
Negative gravity effects  appear related  with a peculiar feature of light surfaces that can be detected by observing photon orbits.
We   investigate  photon circular orbits with orbital frequencies (relativistic velocities) equal in magnitude  to the  \textbf{BH}  horizon frequencies.

The frequency of photon orbits relate \textbf{NSs} to \textbf{BHs}. We shall see that there is a correspondence with singularities having spin mass ratio $M/a$ and $a/M$, known  as\textbf{ BH-NS} dualism--see also\cite{MankoaERuizb}).
 From the observational view point the  detection of these photons, especially close to the  poles ($\theta\approx0$) can  provide information on the property of the horizons  or eventually can  detect a non horizon existence distinguishing \textbf{NSs} from \textbf{BHs},  constituting an observational target  for example for  the Event Horizon Telescope \footnote{\href{}{https://eventhorizontelescope.org/}.}  \cite{ETH1}.
Light surfaces delimiting     stationary observer frequencies in \textbf{NS}  spacetimes   bound  a region,
 Killing throat,  appearing as the ``opening'' and disappearance of  the   Killing horizons of the corresponding \textbf{BH} geometries.
A Killing bottleneck is  a particular structure  of Killing throat  appearing as throat ``restrictions" only   in slowly  spinning \textbf{NSs}.
 Killing bottlenecks appear in association with the concept of  pre-horizon regime introduced in \cite{de-Felice1-frirdtforstati,de-Felice-first-Kerr}. For this reason, Killing bottleneck are also called  horizon remnants.
 There is a pre-horizon regime
in the spacetime  when  there are  mechanical effects  allowing circular orbit  observers
to recognize the close presence of an event horizon. For instance, a gyroscope would observe a ``memory''  of the static  or
stationary initial state
in the Kerr metric. The relevance of this feature   during the gravitational
collapse is discussed in  \cite{de-Felice3,de-Felice-mass,de-Felice4-overspinning,Chakraborty:2016mhx}.
Similar structures     were also studied  in \cite{Tanatarov:2016mcs} and
\cite{Mukherjee:2018cbu,Zaslavskii:2018kix,Zaslavskii:2019pdc}.
It has been argued that
these structures could play an important role
in the description of  \textbf{BH} formation and for testing the possible existence of \textbf{NSs}  \cite{de-Felice3,de-Felice-mass,de-Felice4-overspinning,Chakraborty:2016mhx}.

Here we study Killing bottlenecks, providing  their characteristic frequencies and orbital regions where they can be observed. We frame them in relation to the effects of repulsive gravity and provide a description of the corresponding photon orbits.
Bottleneck structures find  an explanation  in the context of metric bundles (\textbf{MBs}),  which are  particular curves  grouping different  Kerr geometries that allow us to  reinterpret  the concept of horizon and to connect  \textbf{BHs} and \textbf{NSs}  through the photon frequency $\omega$.
The bottleneck is also a characteristic of  the ergoregion.

The effects of repulsive gravity  in the case of the Kerr geometry were  considered in  \cite{1975A&A....45...65D,MankoaERuizb,QM,LQ,Patil:2011aw,Maluf:2014nsa}.
Typical effects of repulsive gravity  were observed in the  naked singularity ergoregion, with  zero and  negative energy states in circular orbits \cite{Stu80,Gariel:2014ara,Pelavas:2000za,Herdeiro:2014jaa,Stuchlik:2014jua}. On the other hand, the presence of negative   energy particles  is a  distinctive feature of the  ergoregion of any spinning source.
Eventually this special matter, constituting an  ``antigravity''
sphere,  can  possibly be filled with negative energy formed according to the Penrose process,
  and bounded by orbits with zero angular momentum (ZAMOS). It has been argued  that
the  ergoregion  cannot  disappear as a consequence of a change in spin, because  it may  be   filled by negative energy matter-- \cite{Chakraborty:2016mhx,Frolov:2014dta} see also \cite{Negative}.
Quantum evaporation of \textbf{NSs} was analyzed  in \cite{Goswami:2005fu}, radiation in \cite{Vaz:1998gd},  and gravitational radiation
in  \cite{Iguchi:1999ud,Iguchi:1998qn,Iguchi:1999jn}\footnote{
An interesting speculative  interpretation of the super-spinner solutions  explores the  duality between elementary particles and \textbf{BHs},
with
quantum \textbf{BH} as the link between
microphysics and macrophysics \cite{Lake:2015pma,Carr:2017grh,Carr:2015nqa,Carr:2014mya,Prok:2008ev}. In any case, the  \textbf{NSs}, which are present in the  Kerr, Reissner-N\"orstrom, and Kerr-Newman spacetimes, provide a perspective of crucial interest regarding the  elementary particles description in general relativity and the definition of the  characteristic  charge, mass, and spin-mass ratio, radius and particles number   of self-gravitating objects\cite{CBS,CBS0,Pu:Charged,Pu:class,Pu:KN,Pu:Kerr,Pu:Neutral}.
In  self-gravitating objects such as  boson stars or fermion stars, the repulsive gravity factors may  enter in the determination of a  critical charge and particle number of the object itself \cite{Carr2020,[18],[19],[20],[21],[22]}. }
Concerning  the ergoregion stability,  see for example, \cite{Cardoso:2007az,CoSch}.)

Finally,  we note that under  general initial data   on the progenitors, the  gravitational collapse analysis  does not rule out   the possibility  that  \textbf{NSs}
can be
produced as the result of a  gravitational collapse \cite{miller,Penrose,Shapiro,Santos,Avi2020,Joshi,Harada,1999JApA...20..233P,Joshi:2001xi,Shapiro-Shapiro,Berger,Bousso}
\footnote{The \textbf{BH}  formation after collapse has been  associated with  trapped surfaces formation, that is, a singularity without trapped surfaces  is usually considered as a proof of its  naked  singularity nature\cite{1999JApA...20..233P,Ziaie:2013tev,Joshi:2001xi,Shapiro-Shapiro,Berger,Bousso}.
Nevertheless, the  non-existence of trapped surfaces after or  during  the gravitational collapse is not a  proof of the existence of a \textbf{NS}.  It is possible to choose a very particular slicing of spacetime during the formation of a spherically symmetric black hole  where  no trapped surfaces exist (see also \cite{Joshi-Book,Wald:1991zz}). }.
 The process of  gravitational collapse   is  constrained by   the Penrose  cosmic
censorship  to  ``good" initial conditions   to end up with horizons formation; on the other hand, the problem of gravitational collapse involves  many factors  including  symmetries during the collapse.
 Consequently, the issue of \textbf{NS} consistency  (\textbf{BH} formation) is still open\cite{Santos,Avi2020,Joshi,Harada}.
Remaining a conjecture, the results depend on the physical initial conditions.
(For  stars or galaxy \textbf{BH} progenitors, with  spin  $a = J/(Mc) $,  usually larger  than the mass $m = GM/c^2$, it is supposed to   lose
mass and   angular momentum during the gravitational collapse towards the singularity.)
The issue of \textbf{NS}  (\textbf{BH}) creation is parallelized with the issue of \textbf{NS} stability--see for example \cite{Sha-teu91,ApTho,J-S09,Jacobson:2010iu,Esitenza,Giacomazzo:2011cv}.

This work is organized as follows.
 Sec.\il(\ref{Sec:setup})   introduces the Kerr geometry and Kerr horizons. Bottlenecks are represented by curves in the  metric bundles  and recognized   by  observing particular photon circular orbits, horizon replicas, which  are defined   in Sec.\il(\ref{Sec:bundle-description}).
 Sec.\il(\ref{Sec:bottlnekc}) contains a  formal definition of bottlenecks,  expressed in terms of the metric bundles and horizon angular momentum  (or photon impact parameter).
Then, bottlenecks are studied on  the  equatorial plane.
A description of the peculiar effects of repulsive gravity on circular motion in the region of the bottleneck is also  provided.
As evidenced by the metric bundles analysis, the bottlenecks are characterized  by the closeness  of the photon  orbital  frequency values  delimiting  also the orbits accessible to the stationary  observers (this range is null on the \textbf{BHs}  horizons).
Each  frequency is demonstrated  to correspond to  a \textbf{BH} horizon frequency.
Bottlenecks can be observed on  horizons replicas, which we discuss in Sec.\il(\ref{Sec:natural-replicas-NS}), considering also the counter-rotating orbits, replicating the frequencies of the inner and outer \textbf{BH} horizons.

\section{Kerr geometry and Killing horizons}
\label{Sec:setup}
The line element  of the  Kerr geometry can be written as
%
\bea\label{Eq:metric-1covector}&&
\dd s^2=-\alpha^2 \dd t^2+\frac{A \sigma}{\rho^2} (\dd \phi- \omega_z \dd t)^2+\frac{\rho^2}{\Delta}\dd r^2+\rho^2 \dd\theta^2,\\&& \sigma\equiv\sin^2\theta,\quad A\equiv (r^2+a^2)^2-a^2 \Delta \sigma\\
&&
\Delta\equiv r^2-2Mr+a^2,\quad\mbox{and}\quad\rho^2\equiv r^2+a^2(1-\sigma) ,
\eea
in Boyer-Lindquist   coordinates.  $M\geq0$  is      interpreted as  the mass  of the gravitational source and  $a\equiv J/M $ is the  specific angular momentum (spin), while   $J$ is the
{total} angular momentum. A naked singularity occurs for $a>M$.
In Eq.\il(\ref{Eq:metric-1covector}),
$\alpha=\sqrt{\Delta \rho^2/A}$  and $\omega_z=2 a M r/A$ are, respectively, the lapse function and the frequency of the zero angular momentum  fiducial observer (ZAMO), whose four-velocity is $u^a=(1/\alpha,0,0,\omega_z/\alpha)$ orthogonal  to the surface   of constant $t$  \cite{observers}.
 In the following, to simplify the discussion, when not otherwise specified, we  use geometrical units with  $r\rightarrow r/M$ and $a\rightarrow a/M$.

\textbf{Killing horizons}

The inner and outer Killing  horizons and the inner and  outer  ergosurfaces  for the {Kerr} geometry  are located at
\bea
r_{\mp}=M\mp\sqrt{M^2-a^2},\quad r_{\epsilon}^{\mp}=M\mp\sqrt{M^2-a^2 \cos ^2\theta},
\eea
respectively.  The  outer  ergosurface, $r_{\epsilon}^+$ is  a \emph{time-like}  surface defined by the vanishing of the norm of  the Killing vector representing  time translations at infinity,  $\partial_t$.
 Killing horizons are the  light-like hypersurface, generated by the flow of a  Killing vector, on
which the norm of the Killing vector vanishes, that is, the  \emph{null} generators coincide with the orbits of an
one-parameter group of isometries: there exists a Killing field $\mathcal{L}$, which is normal to $\mathcal{S}_0$.
The event horizons  of the Kerr  \textbf{BH}  are   Killing horizons   with respect to  the Killing field
$\mathcal{L}_H=\partial_t +\omega_H^{\pm} \partial_{\phi}$, where  $\omega_H^{\pm}$ is the angular velocity {(frequency)} of the horizons, respectively. (The vectors  $\xi_{t}=\partial_{t} $  and
$\xi_{\phi}=\partial_{\phi} $  are, respectively, the stationary
  and axisymmetric  {Killing} fields. In the limiting case of spherically symmetric, static  spacetimes,
 the event horizons are  Killing horizons with {respect} to the  Killing vector
$\partial_t$ and  the
event, apparent, and Killing horizons  with respect to the  Killing field   $\xi_t$ coincide.)
The results we discuss in this work follow from the   investigation of the properties of the Killing vector $\mathcal{L}=\partial_t +\omega \partial_{\phi}$, for photon-like
particles with rotational frequencies $\omega=\omega_{\pm}:\mathcal{L}\cdot \mathcal{L}=0$.
This vector $\mathcal{L}$  is related also to the definition of stationary observes with four-velocity   bounded by the limiting relativistic photon velocities  $ \omega_\pm $. On the \textbf{BH} horizons, $r_{\pm}$, there is $\omega_\pm(r_{\pm})=\omega_H^{\pm}$.
The relation $\mathbf{\mathcal{L_N}}\equiv \mathcal{L}\cdot \mathcal{L}=0$ defines the light-surfaces we consider here.
The ``rotation term'' $\omega_H^+$ plays an important role  in  \textbf{BH} thermodynamics. Indeed,
in the first
first law of \textbf{BH} thermodynamics, for a variation of the total mass $\delta M$,
the term $\omega^+_H (0) \delta J$ can be interpreted as the  ``work term'', where $(0)$ refers to the state prior to the transition)--e.g.
\cite{Wald:1999xu,WWW,inprogress}.
The   Kerr  \textbf{BH} surface gravity  can be written as the combination  $\kappa =\kappa_s-\gamma_a$, where $\kappa_s\equiv { {1}/{4M}}$ is the Schwarzschild surface gravity, while  $\gamma_a=M(\omega_{H}^+)^{2}$ is the contribution due to the additional component of the
\textbf{BH} intrinsic spin. Therefore, $\omega_{H}^+$ is
the  angular velocity, in units of $1/M$, on the event horizon.
In the extreme \textbf{BH} case, the surface gravity  is zero and  the temperature  $T_H = 0$, with consequences also with respect to the stability
 against Hawking radiation. However,   the  entropy (or \textbf{BH} area) of an
extremal \textbf{BH} is not null.
An   implication of the  third law of \textbf{BH} thermodynamics  is  that  there is no physical process reaching from  a non-extremal
\textbf{BH} an   extremal \textbf{BH}  in a finite number of steps \cite{Chrusciel:2012jk,Wald:1999xu,Li:2013sea,Jacobson:2010iu,Sha-teu91,Goswami:2005fu}.
The surface area of the \textbf{BH} event horizon
is non-decreasing in time, which is the content of the second  law.  The thermodynamic laws state also the impossibility  to reach
by a physical
process a \textbf{BH} state with surface gravity $\kappa=0$.

\subsection{Metric bundles and horizons replicas}
\label{Sec:bundle-description}
In \textbf{NSs}, the frequencies $\omega_{\pm}$ are  also  \textbf{BH} horizon frequencies establishing  a \textbf{BH-NS} connection.
The proximity of the frequencies $\omega$ in \textbf{NSs}, with the values of the \textbf{BH} horizons frequencies can be described by  metric bundles (\textbf{MBs}), introduced in \cite{observers,remnants,Pugliese:2019efv,Pugliese:2019rfz,LQG-bundles,PS,bundle-EPJC-complete}.

{\bf Kerr metric bundles:}
A  Kerr metric bundle  is defined as the solution of   $a:\mathcal{L}_{\mathcal{N}}(\bar{\omega})=0$, for a fixed constant  $\bar{\omega}$, (\emph{bundle characteristic frequency}).  It collects  all and only Kerr geometries  with   a (circular) photon orbit (replica) having  the same orbital frequency $\bar{\omega}$, which is the value of  the limiting frequencies $\omega_{\pm}$. A  bundle can be represented by a  curve in the plane   $r/M-\la/M$ ({\emph{extended plane}}), where $\la\equiv a\sqrt{\sigma}$; on the equator the plane is $r-a$.   In  this plane, the curve, $a_{\pm}\equiv \sqrt{r(2M-r)}$,  represents the horizons of all the Kerr \textbf{BHs}. Each point of the curve  is an inner, for $r\in]0,M]$ (with $\omega>1/2$), or outer, for $r\in[M,2M]$ (with $\omega\in ]0,1/2[$), Killing horizon of a Kerr \textbf{BH}--Figs\il(\ref{Fig:featurclosedplb}). The bundle origin,  for $r=0$, is the spin $a_0\equiv1/\omega\sqrt{\sigma}$ (alternatively  $\la_0\equiv1/\omega$).
 All the bundle curves  are  tangent to the horizon curve $a_{\pm}$ and the horizon curve is the envelope surface of all the \textbf{MB} curves. Therefore, the bundle characteristic frequency  $\omega$ is also the   \textbf{BH} horizon frequency  ($\omega_H^+$  or $\omega_H^-$),   defined by the \emph{tangent point} of the bundle curve  to the horizon curve. \textbf{MBs} are represented in Figs\il(\ref{Fig:featurclosedplb}).
  \textbf{MBs} can contain either only \textbf{BH} geometries or \textbf{NS} \textit{and} \textbf{BH} geometries.

\textbf{Horizon replicas and horizon confinement:}
Replicas  connect structures in different  spacetimes  characterized by equal value of the property $\Qa$.
Let $ \Qa ({r_p})$ be  a quantity characteristic of the horizon $r_p$ (for the \textbf{BH} spacetime with spin $a_p$)  in the extended plane.
There is a horizon replica in a spacetime $a_q$, if there exists a radius (orbit)
$r_q$  such that
	$\Qa(r_q,a_q)=\Qa(r_p,a_p)$,  that is, the value $ \Qa({r_p})$  is replicated on a point $r_q$.
In this work,    $\Qa=\omega$.	 All the (\textbf{MB}) frequencies $\omega$  are  \textbf{BH} horizon frequencies;
therefore, here we study horizon  replicas in  \textbf{NS} spacetimes--Sec.\il(\ref{Sec:natural-replicas-NS}).  These structures   reveal a particular  significance in the region proximal  to $(\theta\approx0, \sigma \approx0)$ and frame the bottleneck in the \textbf{MBs}  by the proximity of the bundle curves to the horizon curve in the extended plane, which thickens the curves in the region upper bounded by the inner horizon--Figs\il(\ref{Fig:featurclosedplb}).
There is a  horizon  \emph{confinement}, when   is not possible to find a replica of a value $\Qa$.
In the  Kerr spacetime, part of the inner horizon frequencies are {``confined''},  which means that the horizon frequencies defined for these points of the inner horizons cannot be measured (locally)  outside this region. A part of these frequencies may be detected through measurements close to the \textbf{BH} poles.

\section{Bottlenecks}\label{Sec:bottlnekc}
A bottleneck   is a  restriction of area in the plane $r-\omega$, which is  present in some  naked singularities and is delimited by light surfaces  defined through the condition
$r(\omega):\laa_{\mathcal{N}}(\omega)=0$, i.e., solutions of the form
$r(\omega;a,\theta)$ in terms of the  orbital angular velocity
$\omega$.

\begin{figure*}
      \includegraphics[width=6cm]{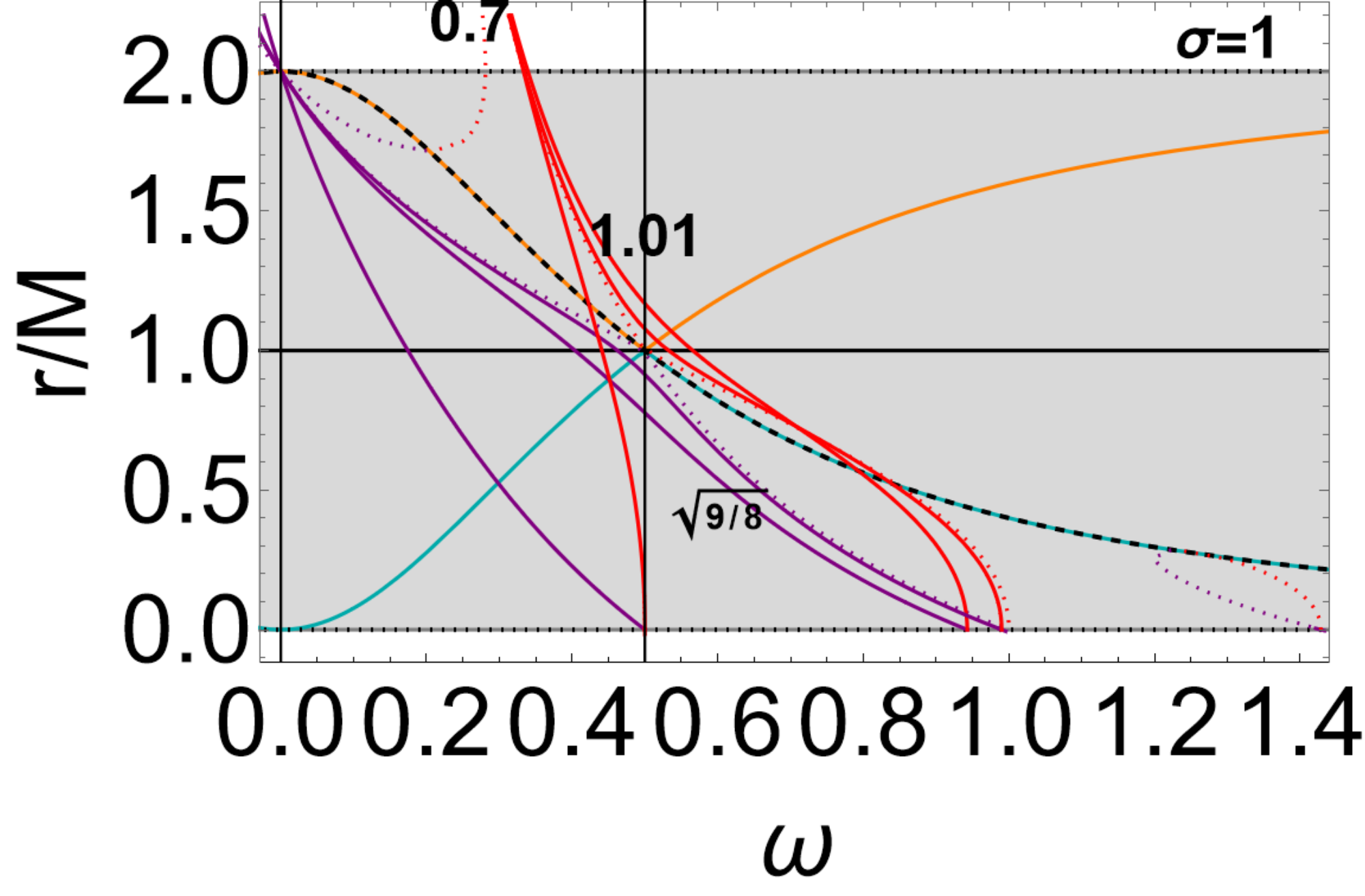}
           \includegraphics[width=6cm]{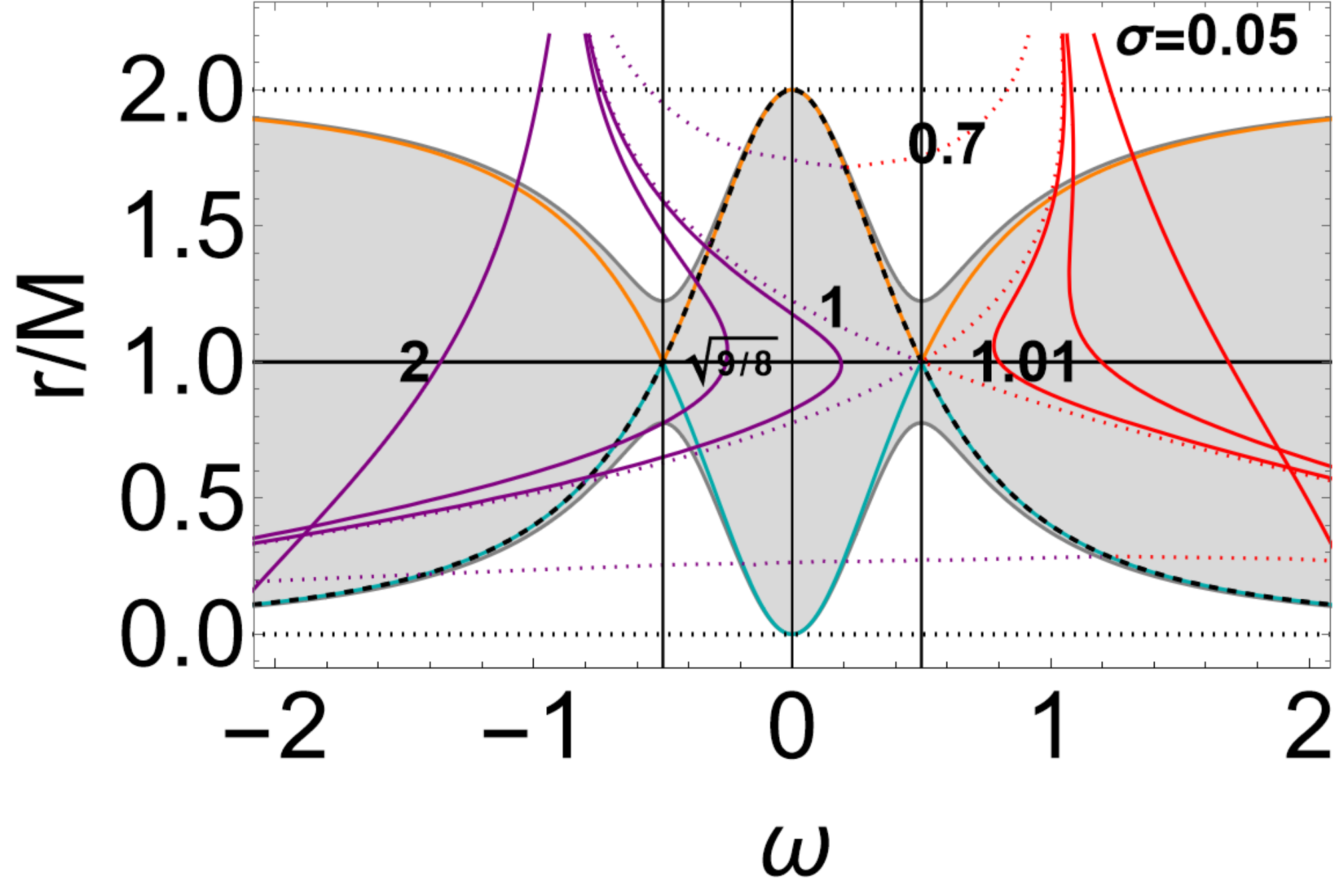}
              \includegraphics[width=6cm]{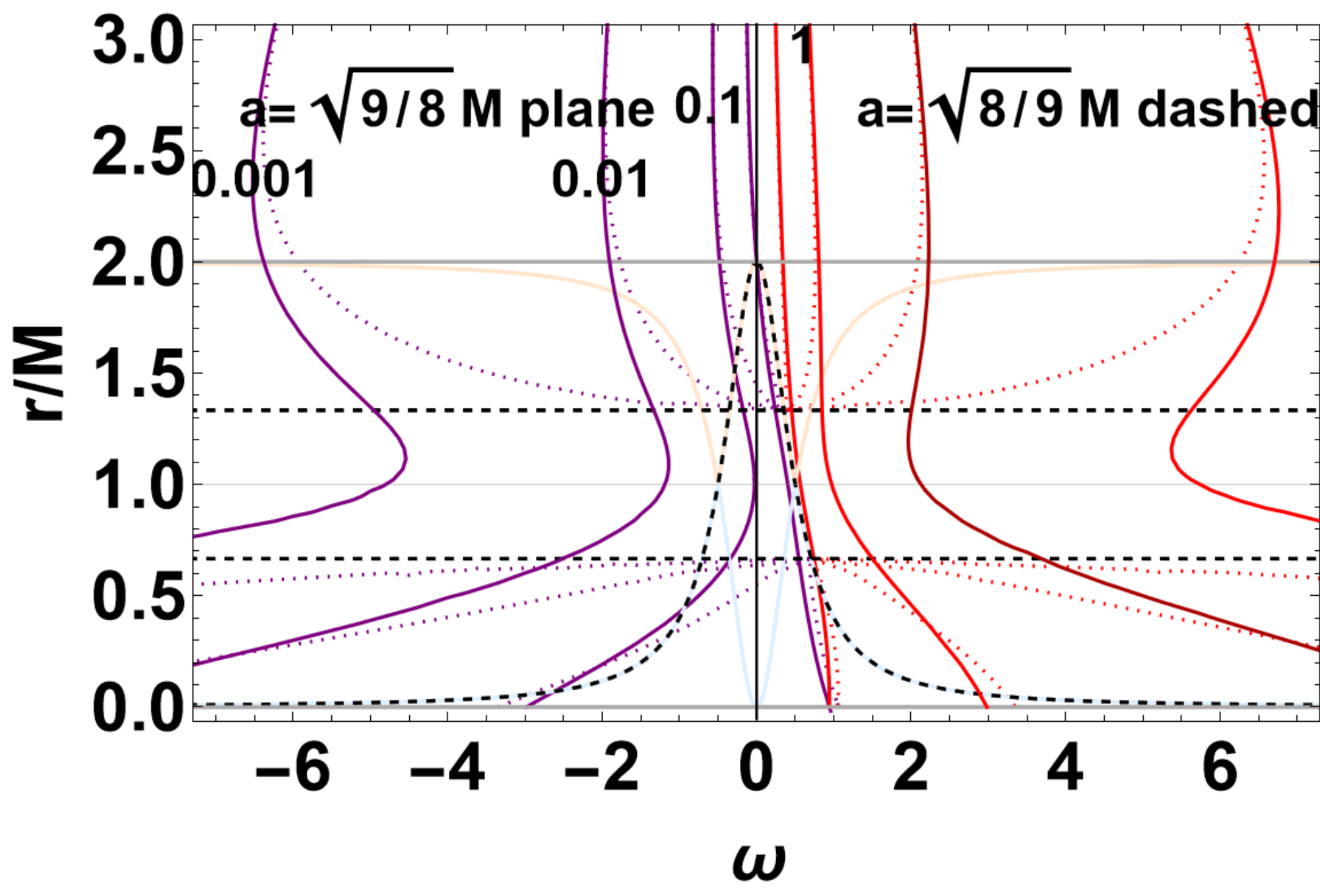}
              \includegraphics[width=6cm]{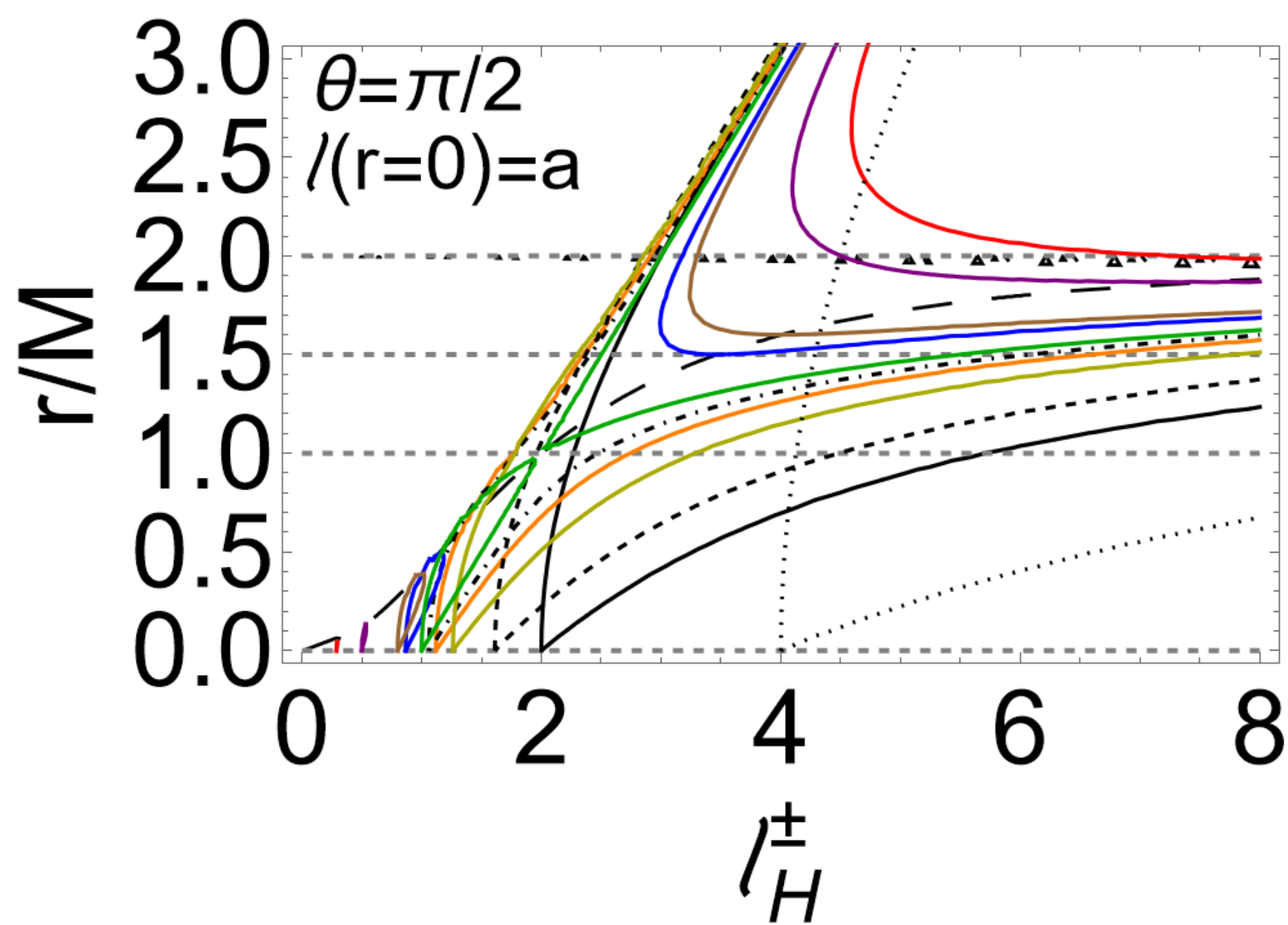}
                \includegraphics[width=6cm]{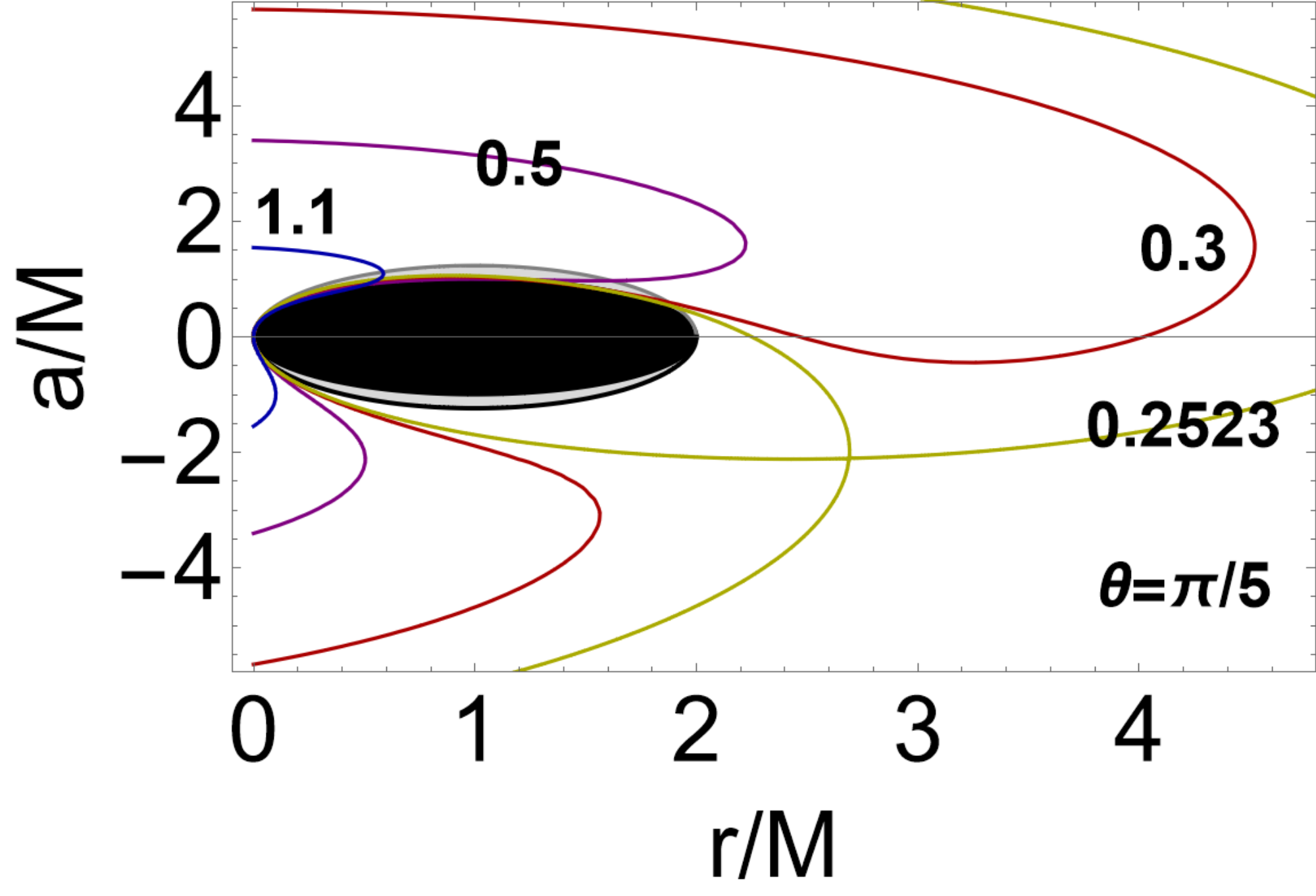}
       \includegraphics[width=6cm]{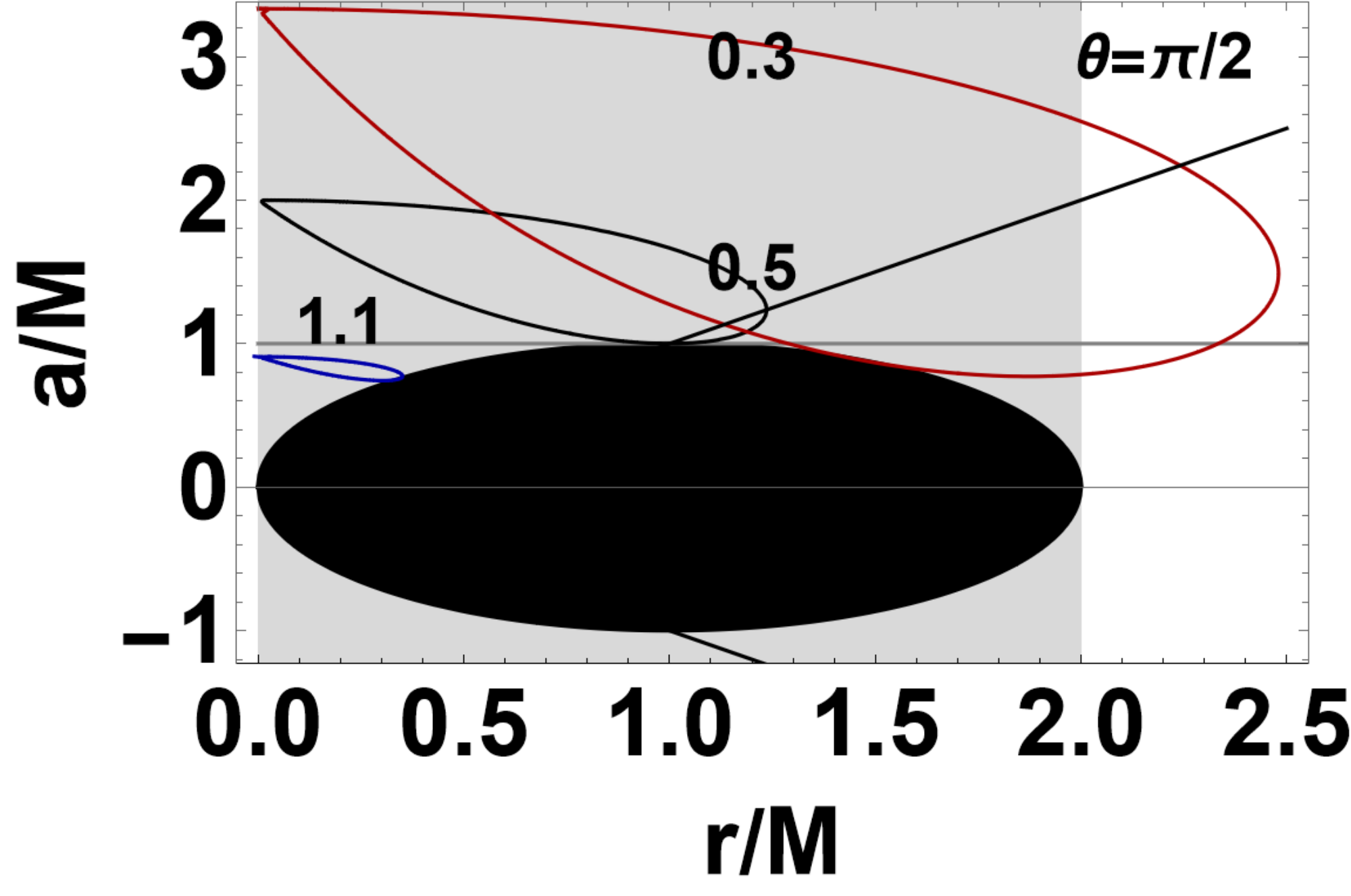}
\caption{In all the panels, gray regions represent the ergoregion $]r_\epsilon^-,r_{\epsilon}^+[$, a black disk corresponds to the central \textbf{BH}  in the extended plane $r/M-a/M$, with boundary $\pm a_{\pm}$. \emph{Bottom line}. Left panel: Light surfaces $r/M$, solutions of $\mathcal{L}\cdot \mathcal{L}=0$ as functions of the specific angular momentum (photon impact parameter) $\ell$.
The functions $\ell_H^{\pm}$ are the outer and inner horizons angular momentum and the bundle origins.  The curves correspond to different spins $a=\ell(r=0)$ on the equatorial plane, including  \textbf{BHs} and \textbf{NSs}. The center and right panels show the metric bundles in the extended plane for selected values of the characteristic frequencies, which are marked on the curves, and for the planes $\theta=\pi/5$ (center) and $\theta=\pi/2$ (right panel).  \emph{Upper line}. Purple  and red curves refers to the bundle frequencies $\omega_-$  and $\omega_+$, respectively. {Orange and  green  curves represent the function $r(\omega)$   horizon curve on the extended plane.}. The black dashed curve corresponds to the \textbf{BH} horizons as  envelope surface of the bundles. Left panel: Light surfaces  $r/M$ on the equatorial plane, i.e., solutions of $\mathcal{L}\cdot \mathcal{L}=0$ as functions of the bundle characteristic (photon) frequencies $\omega$ for different spins, as marked on each curve (see also bottom left panel). The center panel shows the situation on the plane $\sigma=0.6$, where $\sigma\equiv \sin^2\theta$.  Counter-rotating orbits are also shown ($\omega<0$).
Right panel: Light surfaces for different planes $\sigma$ in a \textbf{NS} spacetime with $a/M=\sqrt{9/8}$ (plain curves) and   a \textbf{BH} spacetime with $a/M=\sqrt{8/9}$ (dotted curves). The \textbf{BH} horizons are represented as dashed  horizontal lines.}\label{Fig:featurclosedplb}
 \end{figure*}

 A bottleneck
could be interpreted as due to the existence of two null orbits $r_p\leq r_m$, where $\Delta \omega_{\pm}(r_p,r_m)\equiv \omega^*_+(r_p)-\omega^*_-(r_m)$ is a minimum (or, similarly, the distance  $\Delta r_s^\pm(\omega_p,\omega_m)\equiv r_s^+(\omega_p)-r_s^-(\omega_m)$, where $r_s^{\pm}$ are light surfaces), where $ \omega^*$ is the magnitude of the frequency $\omega$. That is, there exists a pair of points $(r_p,r_m)$ on the  light surfaces of a selected spacetime,
where the interval of photon orbital frequencies  (bounding the range of possible timelike circular orbital frequencies) is minimized.
Clearly,
the limiting case occurs for $r_p=r_m=r_\pm:\Delta \omega_{\pm}(r_\pm)=0$, i.e., on the \textbf{BH} horizons. In this sense,
 the horizons and singularities can be interpreted as  the limiting  cases of the Killing bottlenecks (``horizons remnants"). These structures are shown for different \textbf{NS}  spins $a/M$ and planes $\sigma$ in Figs\il(\ref{Fig:featurclosedplb})
The  bottleneck  is a region approximately around the point $r=M$ and  $\omega=1/2$, that is, the extreme
Kerr  \textbf{BH} in the \textbf{NS} region
 $a\in [M,2M]$  with a further restriction in the range $a\in [M,\sqrt{9/8}M]$, which we explain  in Sec.\il(\ref{Sec:natural-replicas-NS}) through   the analysis of replicas.

 \textbf{Bottleneck and angular momentum}
  Equivalently,  the bottleneck appears  as a region around the point $r=M$ and $ \ell=2$, where
 the   angular momentum (and photon impact parameter) $\ell$ is related to the frequency $\omega$ through
\bea&&\nonumber
\omega \equiv\frac{u^\phi}{u^{t}},\quad 
\ell\equiv\frac{L}{{E}}=
-\frac{u_\phi}{u_t},
\quad
\ell(\omega^{\pm})=\frac{g_{\phi\phi} \omega^\mp}{g_{tt}},\quad \mbox{where}
\\&&
\label{Eq:om-L-far}
\omega_H^\mp=\frac{\sqrt{2-r}}{2 \sqrt{r}},\quad \mbox{and}\quad \ell_H^{\mp}=\frac{2 r}{\sqrt{(2-r)r}}=\la_0=\frac{1}{\omega_H^\mp}.
\eea
%
%
In the Kerr  geometry, the quantities
$
{E} \equiv -g_{\alpha \beta}\xi_{t}^{\alpha} u^{\beta}=-u_t$ and  $ L \equiv
g_{\alpha \beta}\xi_{\phi}^{\alpha}u^{\beta}=u_{\phi}$
are  constants of motion. The constant $L$ may be interpreted       as the axial component of the angular momentum  of a test    particle following
timelike geodesics and $E$ represents the total energy of the test particle
 coming from radial infinity, as measured  by  a static observer at infinity. 
The ``horizons angular momentum'' $\ell_H^{\pm}$ are defined as $\ell(\omega_H^{\pm})$, respectively.  (Note  the limiting condition on the energy process
and $\ell=L/E=\/\omega_H^{\pm}$.).
%
%
Evaluating these quantities on the extended plane, where the horizons are $a_{\pm}=\sqrt{r(2-r)}$ and   $r=r_g$ is a point of the horizon curve ($r\in [0,M]$ for the inner horizons and $r\in[0,2M]$ for the outer horizons, there is   $\omega^{\pm}_H\ell_H^\pm=1$.
Bundles can also be determined by the condition
$\ell=$constant. The relevant aspect is that the angular momentum of the horizon for a fixed frequency is the bundle origin  $\la_0$, giving therefore a direct physical interpretation of the bundle origin.

\textbf{Evaluation of the bottleneck}

\begin{figure}
\centering
   \includegraphics[width=7.81cm]{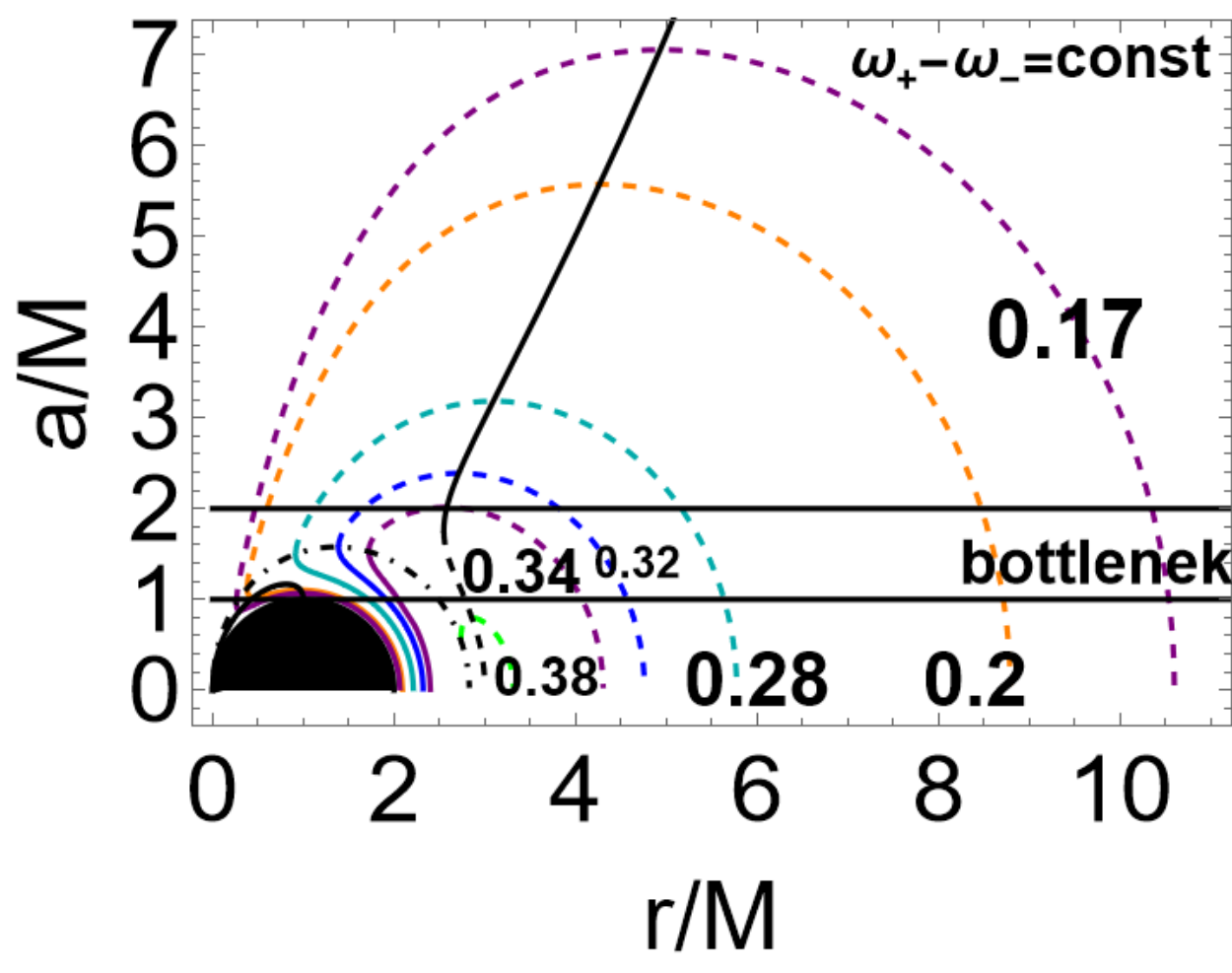}
\includegraphics[width=7.81cm]{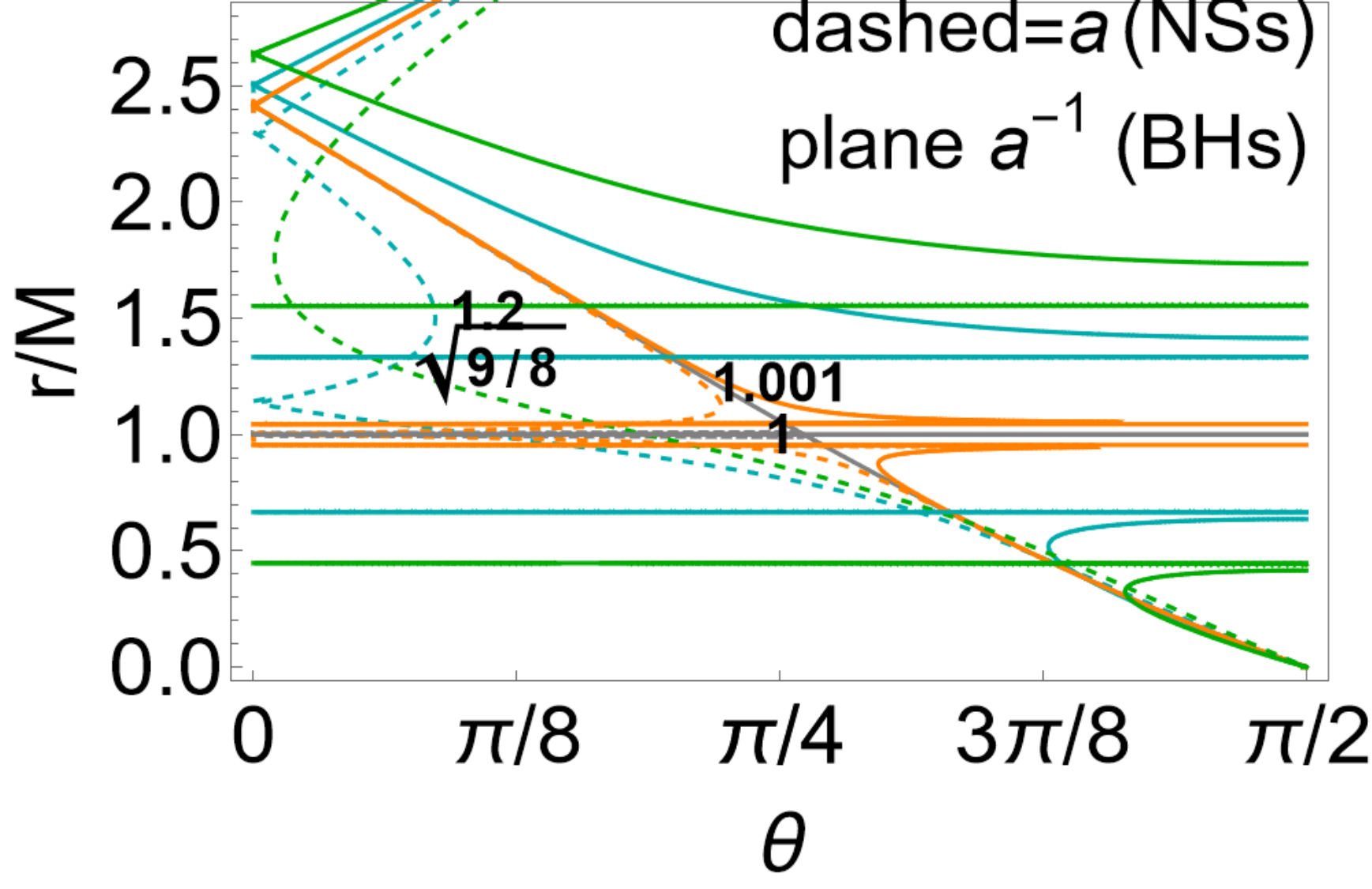}  %
  \caption{Upper panel:  Black region represents black holes in the extended plane.  Curves solutions of frequencies difference   $\Delta \omega_{\pm}\equiv \omega_+-\omega_-=\mathbf{c}=$constant  on the equatorial plane  (curves $a=a_z^{\pm}$) are shown  for selected values of $\mathbf{c}$ marked on the curves.The bottleneck region is highlighted. Black, black dashed, and black dotted-dashed curves are the extreme  $a_\delta^{\pm}:\partial_r \Delta\omega_{\pm}=0$ (plain and dashed respectively)  and  $a_{\delta}^a:\partial_a \Delta\omega_{\pm}=0$  (dotted-dashed curve) of Eqs\il(\ref{Eq:azpm}) {(also radii $r_{\blacksquare}^{\pm}:\partial_{a}\Delta{\omega_{\pm}}=0$  and
 $r_{\Delta}^{\pm}(a):\partial_r\Delta{\omega_{\pm}}=0$
 of Sec.\il(\ref{Sec:bottlnekc}))}.  Bottom panel: Extremes of the photon frequencies (bundles characteristic frequencies) $\omega_{\pm}$ as functions of the radius $r$, on the plane $r/M-\theta$, in the \textbf{NS} geometries with spin $a/M$ (dashed), marked on the curves, and \textbf{BHs} with spins $M/a$ (plain curves).}\label{Fig:Plotsfing1}
\end{figure}

\begin{figure*}
\includegraphics[width=4.45cm]{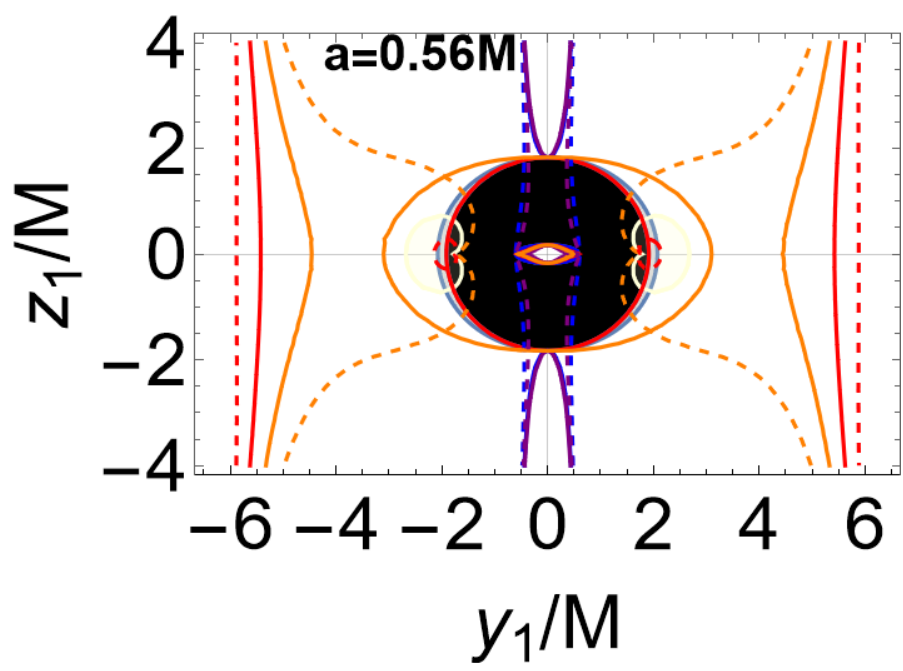}
     \includegraphics[width=4.45cm]{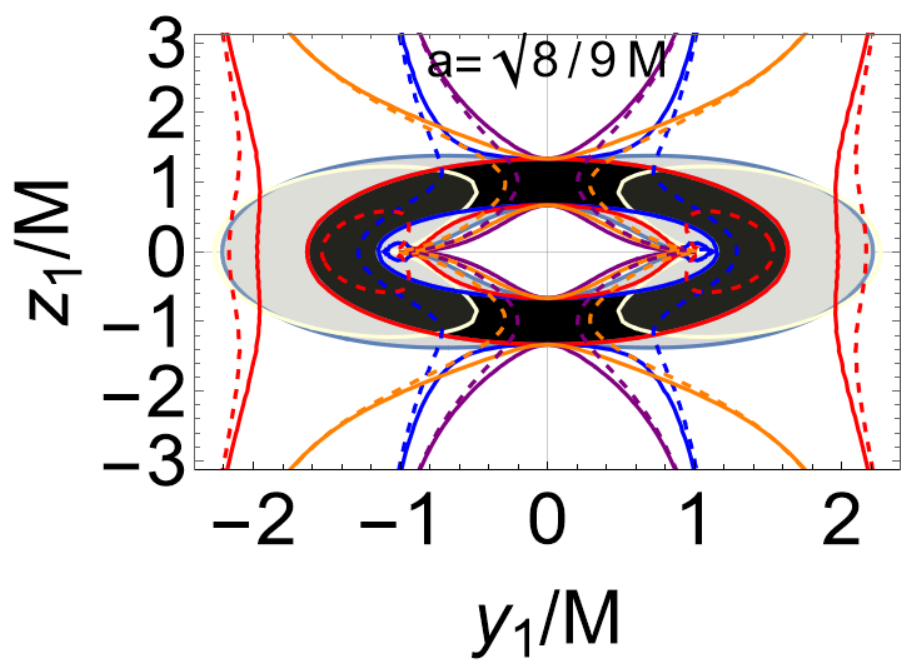}
      \includegraphics[width=4.45cm]{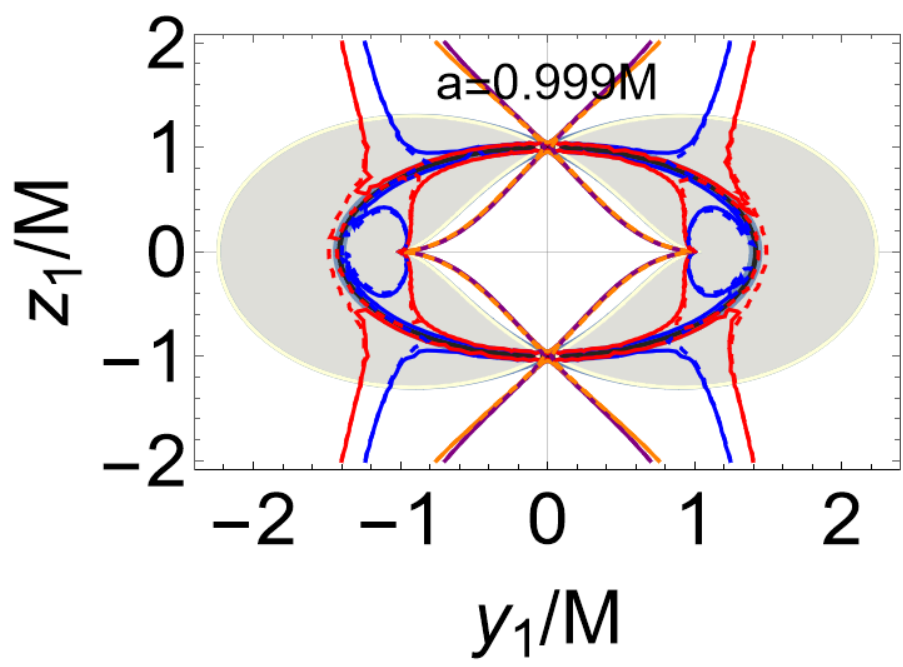}
      \includegraphics[width=4.45cm]{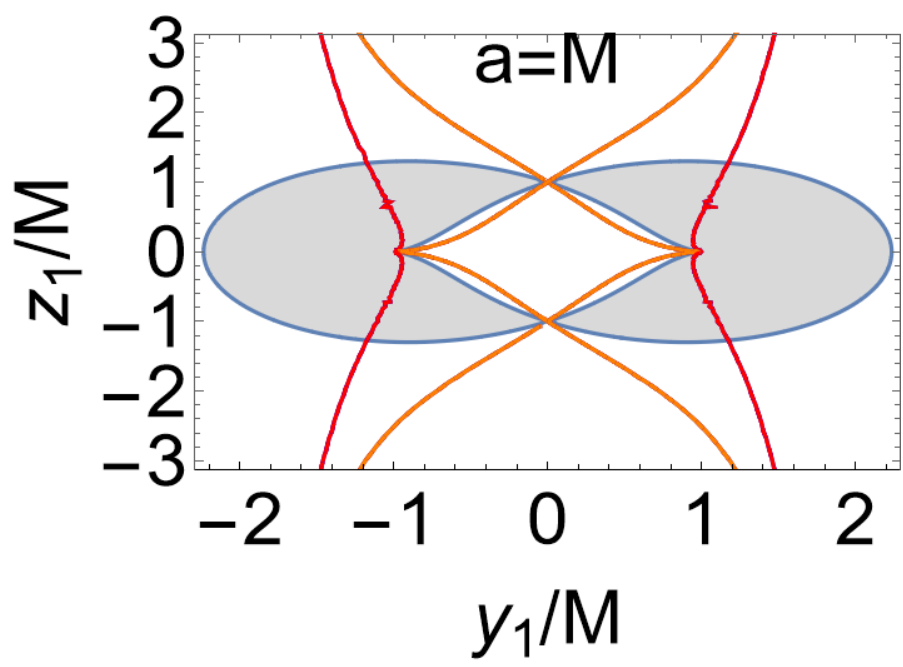}    \\
         \includegraphics[width=19cm]{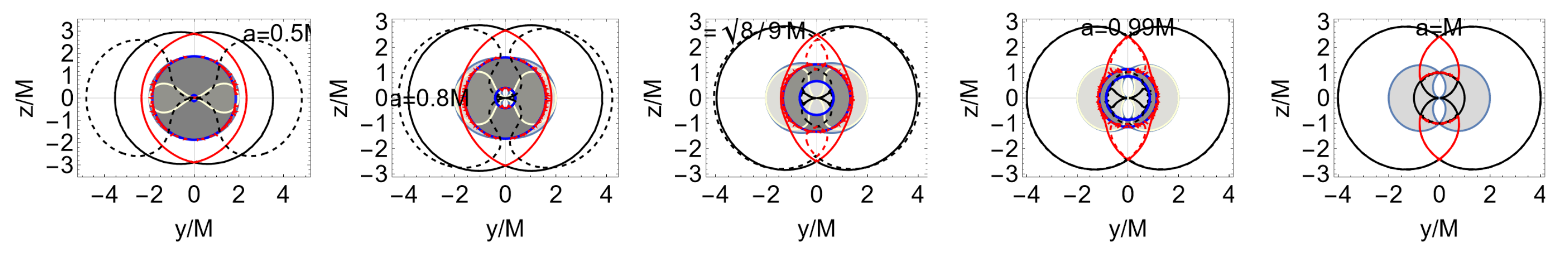}
  \caption{Upper line: Corotating and counter-rotating replicas of the Killing horizons in selected spacetimes: each panel shows the analysis for \textbf{BHs} with dimensionless spin, denoted on the panel, and \textbf{NSs} with spin $M/a$.
  Black region is the  \textbf{BH} $[r_-,r_+]$, where $r_\pm$  are the outer and inner horizons, respectively. Gray region is the ergoregion $]r_{\epsilon}^-,r_{\epsilon}^+[$ in the \textbf{BH} spacetime, where $r_{\epsilon}^{\pm}$ are the outer and inner ergosurface, respectively. Yellow curve represent the radii $r_{\epsilon}^{\pm}$ in \textbf{NSs}. Plane red and  blue curves are corotating  replicas of the inner and outer horizons in the \textbf{BH} spacetime $(\omega_H^\mp)$ respectively. Plane purple and  orange  curves are the counter-rotating   replicas    in the \textbf{BH} spacetime $(-\omega_H^\mp)$. The corresponding replicas in the \textbf{NS} geometry are represented by dashed curves. Standard oblate spheroidal coordinates, $\left\{x_1\equiv\sqrt{a^2+r^2} \sin \theta \cos \phi,y_1\equiv\sqrt{a^2+r^2} \sin\theta\sin\phi,z_1\equiv r \cos\theta\right\}$, have been used. Bottom line: Extremes of the frequencies $\omega_{\pm}$ (red and black curves, respectively) as functions of the radius $r/M$ in selected \textbf{BH} (plain curves) with dimensionless spin $a/M$, denoted on the panel, and \textbf{NS} (dashed curves) with spin $M/a$. The gray region is the range $[r_-,r_+]$. The radii $r_{\pm}$ are blue curves. The light-gray region is the ergoregion $]r_{\epsilon}^-,r_{\epsilon}^+[$ in the \textbf{BH} spacetime. Yellow curves are the  radii $r_{\epsilon}^{\pm}$ in the \textbf{NS} spacetime. There is $ \{x\equiv r \sin\theta \cos \phi,y\equiv r \sin\theta \sin\phi,z\equiv r \cos \theta\}$.}\label{Fig:Plotsfing1m}
\end{figure*}

To analyze  the bottleneck region,  we could  consider the distance
$\Delta \omega_{\pm}=\mathbf{c}$ with  $\mathbf{c}=0$ on the horizon curves $a_{\pm}$.
We evaluate the minimum $\mathbf{c}$ for  \textbf{NSs} by obtaining  the solutions of $(\partial_r \Delta\omega_{\pm}=0)$ (or by minimizing the  two points function $\omega_+(r_p)-\omega_-(r_m)$).
On the equatorial plane, we obtain the solution\footnote{Equation $\Delta \omega_{\pm}=\mathbf{c}$ can be solved for the spins  $a_z^{\pm}\equiv\sqrt{\left[{r^2 \left[2-\mathbf{c}^2 r (r+2)\right]\pm2 \sqrt{r^3 \left[r-4 \mathbf{c}^2 (r+2)\right]}}\right]/{\mathbf{c}^2 (r+2)^2}}$,
where $a_z^-=a_z^+=0$ for  $\mathbf{c}= \pm{2 \sqrt{r-2}}/{r^{3/2}}$. 
Furthermore, it is clear that  the values of $\mathbf{c}$ such that  $a_z^{\pm}=0$,
 which correspond to the Schwarzschild case,  have as extreme case $r=3M$.
 That is, $a_z$   are curves with equal difference in frequency, which is null, as expected, on the horizon in the extended plane- Figs\il(\ref{Fig:Plotsfing1}).}
\bea\nonumber
&&
 a_\delta^{\pm}\equiv \frac{1}{2} \sqrt{r \left[6+(r-3) r\pm\sqrt{(r-1) \left[r^2 (r+3)-36\right]}\right]},
\\\label{Eq:azpm}
&&
\mbox{and there is }\quad
\partial_a \Delta\omega_{\pm}=0\quad \mbox{for}\quad a_{\delta}^a= \sqrt{\frac{r \left(8-r^2\right)}{r+2}}.
\eea
The analysis of the differences  $\Delta \omega_{\pm}$ is indicative of the presence of bottlenecks. The extreme points of the functions $\omega(r)$ are showed also  in Figs\il(\ref{Fig:Plotsfing1}) and  (\ref{Fig:Plotsfing1m}). Figures show the relation between  \textbf{BHs} with spin $a/M$ and \textbf{NSs} with spin $M/a$.  The frequencies extremes  in the bottlenecks are shown in correspondence with the \textbf{BH} horizons.
We note  the emergence in   Figs\il(\ref{Fig:Plotsfing1m}) of the bottleneck region.

\textbf{Effects of repulsive gravity on particles motion}

  The bottleneck region is
  characterized by repulsive gravity effects.
Indeed, on the equatorial plane $(\sigma=1)$, we notice the particular radii
$r_{\blacksquare}^{\pm}:\partial_{a}\Delta{\omega_{\pm}}=0$ {(see corresponding solution $a_{\delta}^{a}$ of Eq.\il(\ref{Eq:azpm})}),
 $r_{\Delta}^{\pm}(a):\partial_r\Delta{\omega_{\pm}}=0$ {(see corresponding solution $a_{\delta}^{\pm}$ of Eq.\il(\ref{Eq:azpm})}), the critical points of the separation $\Delta{\omega_{\pm}}\equiv(\omega_+-\omega_-)$, and the radius $r_{\upsilon}^{\pm}$  with particle orbital energy ${E}=0$, bounding a region filled with  negative  energy orbits. The orbits $\hat{r}_{\pm}$, for which ${L}=0$ (ZAMOs), are also in this region.
The range
  $]\hat{r}_-, \hat{r}_+[$,  is characterized by the presence of counter-rotating circular orbits--see \cite{observers,remnants,ergon,Pugliese:2019efv}.

From  Figs\il(\ref{Fig:featurclosedplb}), the   spin $a_{\infty}=\sqrt{9/8}M$ emerges as a limiting spin for different planes. Near the poles ($\sigma\approx 0$) notice  the closeness of the corotating and counter-rotating frequencies.  The bottleneck exists in a orbital range  centred around $r=M$, corresponding to critical points of the frequencies as functions of $r/M$ (for smaller spins, bottleneck can be studied by analyzing the saddle points), see also Figs\il(\ref{Fig:Plotsfing1}).  The \textbf{NS} geometries  with  spin $a/M>1$, where a bottleneck appears are analysed in correspondence with  \textbf{BHs} of spin $M/a$--Figs\il(\ref{Fig:Plotsfing1}). The closeness of the frequencies in the bottleneck area is also shown in Figs\il(\ref{Fig:Plotsfing1}).

\subsection{Horizon replicas in naked singularities}\label{Sec:natural-replicas-NS}

Let us introduce the following spins and limiting planes $\sigma$:
\bea
&&
a_l\equiv 2,\quad a_s\equiv1.54075, \quad a_t\equiv\sqrt{2},
\\\nonumber
&&a_p\equiv\sqrt{6 \sqrt{3}-9}\equiv1.17996,\quad a_{\infty}\equiv\sqrt{\frac{9}{8}},\quad \sigma_0\equiv \frac{1}{a^2 \omega^2}
\eea
(the limiting plane $\sigma_0$ is related to the bundle origin $a_0\equiv 1/\omega\sqrt{\sigma}$).
There can be one replica, $r_\Qa^{\mathbf{I}}$, or three replicas  $r_\Qa^{\mathbf{III}}$, which are
 solutions of the equation
{\small
\bea&&\nonumber r^4-r^2\frac{a^2 (\sigma -2) \sigma  \omega ^2+1}{ \sigma  \omega ^2}+ 2 r \frac{(a \sigma  \omega -1)^2}{ \sigma  \omega ^2}+
\frac{a^2 (\sigma -1) \left(1-a^2 \sigma  \omega ^2\right)}{ \sigma  \omega ^2}
=0;
\eea}
 in the following ranges of $\sigma$
\bea
&&\mathbf{R_m}\equiv(]0,\sigma _1[\cup] \sigma_2,1[ ,r_\Qa^{\mathbf{I}}), (]\sigma _1,\sigma _2[,r_\Qa^{\mathbf{III}})
\\
&&\mathbf{R_p}\equiv(]0,\sigma _1[\cup] \sigma _2,\sigma_0],r_\Qa^{\mathbf{I}}),(]\sigma _1,\sigma _2[,r_\Qa^{\mathbf{III}})
\\
&&\mathbf{R_x}\equiv(]0,\sigma _1[,r_\Qa^{\mathbf{I}}),(]\sigma _1,1[,r_\Qa^{\mathbf{III}})
\\
&&\mathbf{R_l}\equiv( ]0,\sigma_1[\cup]\sigma_2,\sigma_3[,r_\Qa^{\mathbf{I}}),
(]\sigma_1,\sigma_2[\cup]\sigma_3,1[,r_\Qa^{\mathbf{III}})
\\
&&\mathbf{R_d}\equiv(]\sigma _2,\sigma_0]],r_\Qa^{\mathbf{I}})
\\
&&\mathbf{R_q}\equiv(\sigma\in]0,1[,r_\Qa^{\mathbf{I}}).
\eea
The limiting planes $\sigma_1<\sigma_2<\sigma_3$ are solutions of the equation
 \bea&&
 \sum_{i=0}^6 o_i\sigma^i=0,\quad\mbox{where}
\\\nonumber
&&
o_6\equiv a^6 \omega ^6\left(a^2-1\right)
\\\nonumber
&&
o_5\equiv a^4 \omega ^4 \left[4 a^3 \omega -a^2-\left(a^4+30 a^2-27\right) \omega ^2\right]
\\\nonumber
&&o_4\equiv a^3 \omega ^3 \left[96 a^3 \omega ^3-4 \left(a^4+27\right) \omega ^2+a \left(7 a^2+33\right) \omega -4 a^2\right]
\\\nonumber
&& o_3\equiv  \left[8 a^5 \omega ^3-64 a^6 \omega ^6-6 a^2 \left(a^4+22 a^2-27\right) \omega ^4-6 a^4 \omega ^2\right]
\\\nonumber
&&o_2\equiv  a \omega  \left[96 a^3 \omega ^3-4 \left(a^4+27\right) \omega ^2+a \left(7 a^2+33\right) \omega -4 a^2\right]
\\\nonumber
&&
o_1\equiv  \left[4 a^3 \omega-\left(a^4+30 a^2-27\right) \omega ^2 -a^2\right]
\\\nonumber
&&
o_0\equiv a^2-1.
\eea
The limiting frequencies, $\omega_b$  and $\omega_c$, are  solutions of the equations
\bea\nonumber
&&a \left(a^4+18 a^2-27\right) \omega ^3+3 \left(a^2-3\right)^2 \omega ^2+3 a \left(a^2-3\right) \omega +a^2=0;
\\
&& a \left(a^2+27\right) \omega ^3+\left(3 a^2-27\right) \omega ^2+3 a \omega +1=0;
\eea
respectively.
Note that the equatorial plane, $\sigma=1$, the geometry $a=a_{\infty}$, and the extreme \textbf{BH}  frequency $\omega=1/2$, are special cases.
Below we provide the complete prospectus  of the  corotating  $\omega>0$  and counter-rotating $\omega<0$ replicas of the inner and outer horizons in the extended plane,  detectable in the \textbf{NS} geometries.

\textbf{Corotating  inner horizon replicas}
\bea&&\nonumber
a\in ]1,a_{\infty}[:\left(\omega\in \left]\frac{1}{2},\omega_b\right[,\mathbf{R_x}\right),\left(\omega\in\left]\omega_b,\frac{1}{a}\right[,\mathbf{R_l}\right),\left(\omega >\frac{1}{a},\mathbf{R_p}\right)
\\\nonumber
&&a\in ]a_{\infty },a_p[:\left(\omega\in \left]\frac{1}{2},\frac{1}{a}\right[,\mathbf{R_x}\right),\left(\omega\in\left]\frac{1}{a}, \omega_b\right],\mathbf{R_d}\right),(\omega >\omega_b,\mathbf{R_p});
\\\nonumber
&&a\in [a_p,a_l[:\left(\omega\in\left]\frac{1}{2},\frac{1}{a}\right[,\mathbf{R_x}\right),\left(\omega >\frac{1}{a},\mathbf{R_d}\right);
\\
&&a\geq a_l:\left(\omega >\frac{1}{2},\mathbf{R_d}\right).
\eea

\textbf{Corotating outer horizon replicas}
\bea&&\nonumber
a\in ]1, a_l]:\left(\omega \in \left]0,\frac{1}{2}\right[,\mathbf{R_x}\right);
\\
&& a>a_l:\left(\omega\in\left ]0,\frac{1}{a}\right[,\mathbf{R_x}\right),\left(\omega\in\left ]\frac{1}{a},\frac{1}{2}\right[,\mathbf{R_d}\right).
\eea
\textbf{Counter-rotating outer horizon replicas}
\bea&&
a\in]1,a_s]:\left(\omega\in\left]-\frac{1}{2},\omega_c\right[,\mathbf{R_m}\right),(\omega\in]\omega_c, 0[,\mathbf{R_x});
\\\nonumber
&&
a\in]a_s,a_l]:\left(\omega\in\left]-\frac{1}{2}, \omega_b\right],\mathbf{R_q}\right),(\omega\in]\omega_b,\omega_c[,\mathbf{R_m}),
\\
&&\nonumber
\qquad\qquad\quad(\omega\in]\omega_c,0[,\mathbf{R_x});
\\\nonumber
&&a>a_l:\left(\omega\in\left]-\frac{1}{2},-\frac{1}{a}\right[,\mathbf{R_d}\right),\left(\omega\in\left[-\frac{1}{a}, \omega_b\right],\mathbf{R_q}\right),
\\
&&\nonumber\qquad \quad (\omega\in]\omega_b,\omega_c[,\mathbf{R_m}),(\omega\in]\omega_c, 0[,\mathbf{R_x});
\eea
\textbf{Counter-rotating inner horizon replicas}
{\small
\bea&&\nonumber
 a\in ]1,a_p]:\left(\omega <-\frac{1}{a},\mathbf{R_p}\right),\left(\omega\in \left]-\frac{1}{a},-\frac{1}{2}\right[,\mathbf{R_m}\right);
 \\\nonumber
 && a\in]a_p,a_t[:(\omega \leq \omega_b,\mathbf{R_d}),\left(\omega\in\left]\omega_b,-\frac{1}{a}\right[,\mathbf{R_p}\right),\left(\omega\in\left[-\frac{1}{a},-\frac{1}{2}\right[,
 \mathbf{R_m}\right);
 \\\nonumber
 &&a\in ]a_t,a_s[:\left(\omega <-\frac{1}{a},\mathbf{R_d}\right),\left(\omega\in\left[-\frac{1}{a}, \omega_b\right],\mathbf{R_q}\right),\left(\omega\in\left]\omega_b,-\frac{1}{2}\right[,\mathbf{R_m}\right);
 \\\nonumber
 &&a\in [a_s,a_l]:\left(\omega <-\frac{1}{a},\mathbf{R_d}\right),\left(\omega\in\left[-\frac{1}{a},-\frac{1}{2}\right[,\mathbf{R_q}\right);
 \\
 &&a>a_l:\left(\omega <-\frac{1}{2},\mathbf{R_d}\right).
\eea}
 Examples of replicas are shown in Figs\il(\ref{Fig:Plotsfing1m}).  Replicas ranges are more articulated, as expected,  in the counter-rotating case and for the inner horizon replicas. There is a limiting frequency $1/a$, which is related to the bundle origin and the horizon angular momentum $\ell$. Furthermore, it frames the correspondence between \textbf{NSs} with spins $a/M$ and \textbf{NSs} with spin $M/a$.

\section{Conclusion}
In this work, we presented   evidences of repulsive gravity effects  in Kerr naked singularities by analyzing photon orbits, replicas and  bottlenecks--Sec.\il(\ref{Sec:natural-replicas-NS}).
Replicas, photon orbits with frequency $\omega$  (bundles characteristic frequency) equal  in magnitude with the  \textbf{BHs} inner and outer horizons, can extract ``information" on the inner horizon frequencies  near the poles ($\sigma\approx 0$):
close to the rotational axis $\sigma\ll1$,  for a spacetime $a$ it is possible to find an inner horizon replica  in the outer region  $r>r_+$. We  completed the analysis of the  corotating and counter-rotating  (inner and outer) horizons replicas. We showed  the correspondence between \textbf{BH} and \textbf{NS}  through the frequencies $\omega$, framed in the  metric bundles approach of Sec.\il(\ref{Sec:bottlnekc})--Figs\il(\ref{Fig:featurclosedplb}).
Bottlenecks appear as a restriction of the area of the plane  $r-\omega$ or  $r-\ell$ (angular momentum of the horizon and photon impact parameter)  defined by  light surfaces,
 as  a characteristic of slowly spinning \textbf{NSs}, which in the \textbf{BH} case coincide with the Killing horizons; therefore, bottlenecks are also defined as horizon remnants.

The bottleneck analysis shows also the correspondence between \textbf{NSs} with dimensionless spin  $a/M$ and \textbf{BHs} with spin $M/a$, a \textbf{ BH-NS} dualism. The ratio is related to the bundle frequency and angular momentum--Figs\il(\ref{Fig:Plotsfing1m}).

\section*{Acknowledgements}
This work was partially supported  by UNAM-DGAPA-PAPIIT, Grant No. 114520,
Conacyt-Mexico, Grant No. A1-S-31269,
and by the Ministry of Education and Science  of Kazakhstan, Grant No.
BR05236730 and AP05133630.


\begin{thebibliography}{99}
\bibitem
{bundle-EPJC-complete}
 D. Pugliese\&H. Quevedo,
 Eur. Phys. J. \textbf{C 81}  3 (2021) 258.




 \bibitem{observers}
D.~Pugliese\&H.~Quevedo,
  Eur.\ Phys.\ J.\ C {\bf 78} 1  (2018) 69.

  \bibitem{remnants} D.~Pugliese\&H.~Quevedo,
  Eur.\ Phys.\ J.\ C {\bf 79} 3 (2019) 209.


\bibitem{Pugliese:2019efv}
  D.~Pugliese\&H.~Quevedo,
  arXiv:1910.02808 [gr-qc] (2019).
\bibitem{Pugliese:2019rfz}
  D.~Pugliese\&H.~Quevedo,
  arXiv:1910.04996 [gr-qc] (2019).

 \bibitem{LQG-bundles}
D.~Pugliese\& G.~Montani,
Entropy \textbf{22}   (2020) 402.
\bibitem{PS} D.  Pugliese \& Z. Stuchlik, Class.Quant.Grav. https://doi.org/10.1088/1361-6382/abff97 (2021).


\bibitem{MankoaERuizb}
V. S. Manko\&E. Ruiz,
Phys. Lett. B. \textbf{791}  (2019)  26--29.



\bibitem{ETH1}  The Event Horizon Telescope Collaboration et al.,  ApJL \textbf{875} (2019)  L1.


 \bibitem{de-Felice1-frirdtforstati}
  F. de Felice,
Mont. Notice R. astr. Soc  \textbf{252}  (1991)  197--202.

\bibitem{de-Felice-first-Kerr}
 F.  de Felice and  S. Usseglio-Tomasset, Class.Quant.Grav. \textbf{8}  (1991)  1871--1880.




\bibitem{de-Felice3}  F.  de Felice\&S. Usseglio-Tomasset, Gen.Rel.Grav.  \textbf{24} (1992) 10.


\bibitem{de-Felice-mass}  F.  de Felice and  Y. Yunqiang, Class.Quant.Grav. \textbf{10}  (1993) 353--364.


\bibitem{de-Felice4-overspinning}  F.  de Felice  and  L. Di G. Sigalotti, Ap.J. \textbf{389} (1992) 386--391.


 \bibitem{Chakraborty:2016mhx}
   C. Chakraborty, M. Patil, et al., 
  Phys.Rev.D {\bf 95} 8  (2017) 084024.


\bibitem{Tanatarov:2016mcs}
  I.~V.  Tanatarov and O.~B. Zaslavskii,
  Gen.Rel.Grav.  {\bf 49} 9  (2017)  119.


\bibitem{Mukherjee:2018cbu}
  S. Mukherjee and R.~K. Nayak,
  Astrophys.\ Space Sci.\  {\bf 363} 8 (2018)  163.


\bibitem{Zaslavskii:2018kix}
  O.~B Zaslavskii.,
  Phys.Rev.D {\bf 98} 10  (2018)  104030.

\bibitem{Zaslavskii:2019pdc}
  O.~B. Zaslavskii,
  Phys.Rev.D {\bf 100}  2 (2019)  024050.

\bibitem{1975A&A....45...65D} F. de Felice, \aap \textbf{45} (1975) 65.

\bibitem{QM}
D. Batic, D. Chin,   M.  Nowakowski, 
 Eur.Phys.J.C \textbf{71}  (2011) 1624.
\bibitem{LQ}
O. Luongo\&H. Quevedo,
Phys.Rev.D \textbf{90} (2014) 084032.
\bibitem{Patil:2011aw}
M.~Patil, P.~S.~Joshi and D.~Malafarina,
Phys.Rev.D {\textbf{83}} (2011) 064007.
\bibitem{Maluf:2014nsa}
J.~W.~Maluf,
Gen. Rel. Grav. {\textbf{46}} (2014) 1734.


 \bibitem{Stu80}
Z. Stuchlik,
Bull. Astron. Inst. Czech  \textbf{31} (1980)   129.

\bibitem{Gariel:2014ara}
  J.~Gariel, N.~O.~Santos and J.~Silk,
  Phys.Rev.D {\bf 90} (2014) 063505.

  \bibitem{Pelavas:2000za}
  N.~Pelavas, N.~Neary and K.~Lake,
  Class.Quant.Grav.  {\bf 18}  (2001) 1319.

  \bibitem{Herdeiro:2014jaa}
  C.~Herdeiro and E.~Radu,
  Phys.Rev.D {\bf 89} (2014) 1240 18.


  \bibitem{Stuchlik:2014jua}
   Z. Stuchlik,  D.  Pugliese, J.  Schee \& H. Kucáková,
  Eur.\ Phys.\ J.\ \textbf{C { 75}}  9  (2015)  451.


\bibitem{Frolov:2014dta}
  A.~V.~Frolov and V.~P.~Frolov,
  Phys.Rev.D {\bf 90} 12 (2014)  124010.


\bibitem{Negative}
M. B. Paranjape, Physics Today,  in Commentary\&Reviews 24 May 2017.

\bibitem{Goswami:2005fu}
  R.~Goswami, P.~S.~Joshi and P.~Singh,
  Phys.Rev.Lett.   {\bf 96}   (2006) 031302.

\bibitem{Vaz:1998gd}
  C.~Vaz\&L.~Witten,
  Phys.Lett.B {\bf 442}   (1998)  90.


\bibitem{Iguchi:1999ud}
  H.~Iguchi, T.~Harada and K.~i.~Nakao,
  Prog.\ Theor.\ Phys.\  {\bf 101}  (1999)   1235.
\bibitem{Iguchi:1998qn}
  H.~Iguchi, K.~i.~Nakao and T.~Harada,
  Phys.Rev.D {\bf 57}  (1998)  7262.

\bibitem{Iguchi:1999jn}
  H.~Iguchi, T.~Harada and K.~I.~Nakao,
  Prog.\ Theor.\ Phys.\  {\bf 103} (2000) 53.




\bibitem{Lake:2015pma}
  M.~J.~Lake\&B.~Carr,
  JHEP {\bf 1511}   (2015) 105.


\bibitem{Carr:2017grh}
  B.~J.~Carr,
  arXiv:1703.08655 [gr-qc].

\bibitem{Carr:2015nqa}
  B.~J.~Carr, J.~Mureika and P.~Nicolini,
  JHEP {\bf 1507}  (2015)  052.

  \bibitem{Carr:2014mya}
  B.~J.~Carr,
  Springer Proc.\ Phys.\  {\bf 170}   (2016) 159.


\bibitem{Prok:2008ev}
  Y.~Prok {\it et al.} [CLAS Collaboration],
  Phys.Lett.B {\bf 672} (2009) 12.

\bibitem{CBS}
D.~Pugliese, H.~Quevedo, J.~A.~Rueda H. and R.~Ruffini,
Phys.Rev.D \textbf{88} (2013) 024053.
\bibitem{CBS0} R. Ruffini and S. Bonazzola, Phys. Rev. \textbf{187}  (1969) 1767.




\bibitem{Pu:Charged}
 D.  Pugliese,  H. Quevedo and R.  Ruffini,
  Phys.Rev.D {\bf 83} (2011) 104052.

\bibitem{Pu:class}
  D.  Pugliese,  H. Quevedo and R.  Ruffini,
  Eur.\ Phys.\ J.\ C {\bf 77} 4   (2017) 206.

\bibitem{Pu:KN}
  D.  Pugliese,  H. Quevedo and R.  Ruffini,
  Phys.Rev.D {\bf 88} (2013)  024042.





\bibitem
  {Pu:Kerr}
D. Pugliese,  H. Quevedo and R. Ruffini,
  \prd {84} (2011) 044030.
\bibitem{Pu:Neutral}
 D. Pugliese, H. Quevedo and R.  Ruffini,
  Phys.\ Rev.\  D {\bf 83} (2011) 024021.

\bibitem{Carr2020} B. Carr, H. Mentzer, J. Mureika, P. Nicolini, Eur.Phys.J. C   \textbf{80} (2020) 1166.
\bibitem{[18]}C. Sivaram and K. P. Sinha, Phys. Rev. \textbf{D 16} (1977) 1975.
\bibitem{[19]}A. Salam and J.A. Strathdee, Phys. Rev. \textbf{D 18}  (1978) 4596.
\bibitem{[20]} C. F. E. Holzhey and F. Wilczek, Nucl. Phys. \textbf{B 380}  (1992) 447.
\bibitem{[21]}R.L. Oldershaw and  J. Cosmol. 6 (2010) 1361.
\bibitem{[22]}B.J. Carr, Mod. Phys. Lett.\textbf{ A 28} (2013) 1340011.

 \bibitem{Cardoso:2007az}
  V.~Cardoso, P.~Pani, M.~Cadoni and M.~Cavaglia,
  Phys.Rev.D {\bf 77}  (2008) 124 044.

\bibitem{CoSch}
N. Comins\& B. F. Schutz,
Proc.  R. Soc. A
 \textbf{364}  1717    (1978) 211--226.


\bibitem{miller}
  A.~Helou, I.~Musco, J.~C.~Miller,
  Class.Quant.Grav.  {\bf 34}  13  (2017) 135012.


   \bibitem{Penrose}
    R. Penrose, Nuovo Cimento Rivista Serie (1969) 1.


\bibitem{Shapiro}
S. L. Shapiro\& S. A. Teukolsky,
American Scientist
\textbf{79}  4 (1991)  330--343.
\bibitem{Santos}
T.  Crisford and J. E. Santos,
Phys. Rev. Lett. \textbf{118} (2017) 181101.

\bibitem{Avi2020}
A. Loeb, \emph{In Search of Naked Singularities}
 Scientific American Observations-Opinion May 3 (2020).

\bibitem{Joshi}
P. S. Joshi,
Scientific American
\textbf{300} 2 (2009)  36--43.


\bibitem{Harada}
T. Harada,
Pramana \textbf{63}  4
 (2004)
741--753.


\bibitem
{1999JApA...20..233P} R.  Penrose, JJApA 20 (1999) 233.

\bibitem{Joshi:2001xi}
P.~S.~Joshi, N.~Dadhich\& R.~Maartens,
Phys.Rev.D \textbf{65} (2002) 101501.

\bibitem{Shapiro-Shapiro}
S. L. Shapiro and S. A. Teukolsky, 
Philosophical Transactions: Physical Sciences and Engineering  \textbf{340}  1658 (1992) 365--390.



\bibitem{Berger}
B. K.  Berger, 
Living Reviews in Relativity \textbf{5} 1 (2002) 1.

\bibitem{Bousso}
R. Bousso, A. Shahbazi--Moghaddam,  M. Tomasevic,
Phys.Rev.D \textbf{100} (2019) 126010.

\bibitem{Ziaie:2013tev}
A.~H.~Ziaie, A.~Ranjbar\& H.~R.~Sepangi,
Class. Quant. Grav. \textbf{32} (2015) 025010.



 \bibitem{Joshi-Book}  P. S. Joshi,  \emph{Gravitational Collapse and Spacetime Singularities},
 Cambridge Monographs on Mathematical Physics, New York (2007).

 \bibitem{Wald:1991zz}
  R.~M.~Wald and V.~Iyer,
  Phys.Rev.D {\bf 44}  (1991) 3719.


\bibitem{Sha-teu91}
 S. L. Shapiro\& S. A. Teukolsky,
Phys. Rev. Lett. \textbf{66}  (1991) 994.

\bibitem{ApTho}
T. A. Apostolatos\& K. S.  Thorne,
 Phys. Rev. \textbf{D} \textbf{46} (1992) 2435.



\bibitem{J-S09}  T. Jacobson\& T. P. Sotiriou, Phys. Rev. Lett. \textbf{103}
(2009) 141101.

\bibitem{Jacobson:2010iu}
  T.~Jacobson\& T.~P.~Sotiriou,
  J. Phys.  Conf. Ser.   {\bf 222}  (2010) 012041.

  \bibitem{Esitenza}
E. Barausse, V. Cardoso,  G. Khanna,
Phys.Rev.Lett.\textbf{105}  (2010) 261102.

\bibitem{Giacomazzo:2011cv}
  B.~Giacomazzo, L.~Rezzolla, N.~Stergioulas,
  Phys.Rev.D {\bf 84}  (2011) 024022.


\bibitem{Wald:1999xu}
 R.~M. Wald,
  Class.Quant.Grav.  {\bf 16} (1999) A177.

\bibitem{WWW} R. M. Wald,
Living Rev. Relativ.  4(1) (2001) 6.
 \bibitem{inprogress} D.~Pugliese and H.~Quevedo, 2021,  {to be submitted}


\bibitem{Chrusciel:2012jk}
  P. T. Chrusciel, J.Lopes Costa\& M. Heusler,
  Living Rev. Rel. {\bf 15}  (2012) 7.


\bibitem{Li:2013sea}
  Z.~Li and C.~Bambi,
  Phys.Rev.D {\bf 87},   124022 (2013).


\bibitem{ergon}
D.~Pugliese and H.~Quevedo,
  Eur.\ Phys.\ J.\ C {\bf 75} 5  (2015) 234.

\end{thebibliography}
\end{document}